\documentstyle[12pt,epsfig]{article}
\begin{document}
\begin{center}
{\bf ELECTROMAGNETIC PROPERTIES OF THE SU(3) OCTET BARYONS     \\
IN THE SEMIBOSONIZED SU(3) NAMBU--JONA-LASINIO MODEL}
\footnote{Talk given at 7th International Conference on the Structure of
Baryons, Santa Fe, New Mexico, 3-7 Oct 1995.} \\
\vspace{5mm}
HYUN-CHUL KIM, ANDREE BLOTZ
\footnote{Present address: Department of Physics, State University
of New York, Stony Brook, 11794, U.S.A.},
MAXIM V. POLYAKOV
\footnote{On leave of absence from Petersburg Nuclear Physics
Institute, Gatchina, St. Petersburg 188350, Russia}
and KLAUS GOEKE \\
\vspace{5mm}
{\it Institute for  Theoretical  Physics  II, \\  P.O. Box 102148,
Ruhr-University Bochum, \\
D--44780 Bochum, Germany
       } \\
\vspace{5mm}
\end{center}
\date{December 1995 }
\begin{abstract}
The electromagnetic properties of the SU(3) octet baryons
are investigated in the semibosonized SU(3) Nambu--Jona-Lasinio
model.The rotational $1/N_c$ and strange quark mass corrections
in linear order are taken into account.
It turns out that the model is in good agreement with
the experimental data.
\end{abstract}
\vspace{5mm}
The baryon in the semibosonized Nambu--Jona-Lasinio model
(often called chiral quark-soliton model) is
regarded as $N_c$ valence quarks coupled to the polarized
Dirac sea bound by a non-trivial chiral field configuration
in the Hartree approximation~\cite{dpp}.  This picture of the
baryon can be justified in the large $N_c$ limit\cite{Witten}.
The model has been found
to be very successful in describing the static properties of the
nucleon and its form factors\cite{getal}--\cite{chetal} in SU(2).
The model was generalized to SU(3) by Weigel {\em et al.}~\cite{war}
and Blotz {\em et al.}\cite{betal2} and the mass-splittings of hyperons
were in remarkable agreement with experimental data.

\vspace{0.8cm}
\centerline{\epsfysize=3.2in\epsffile{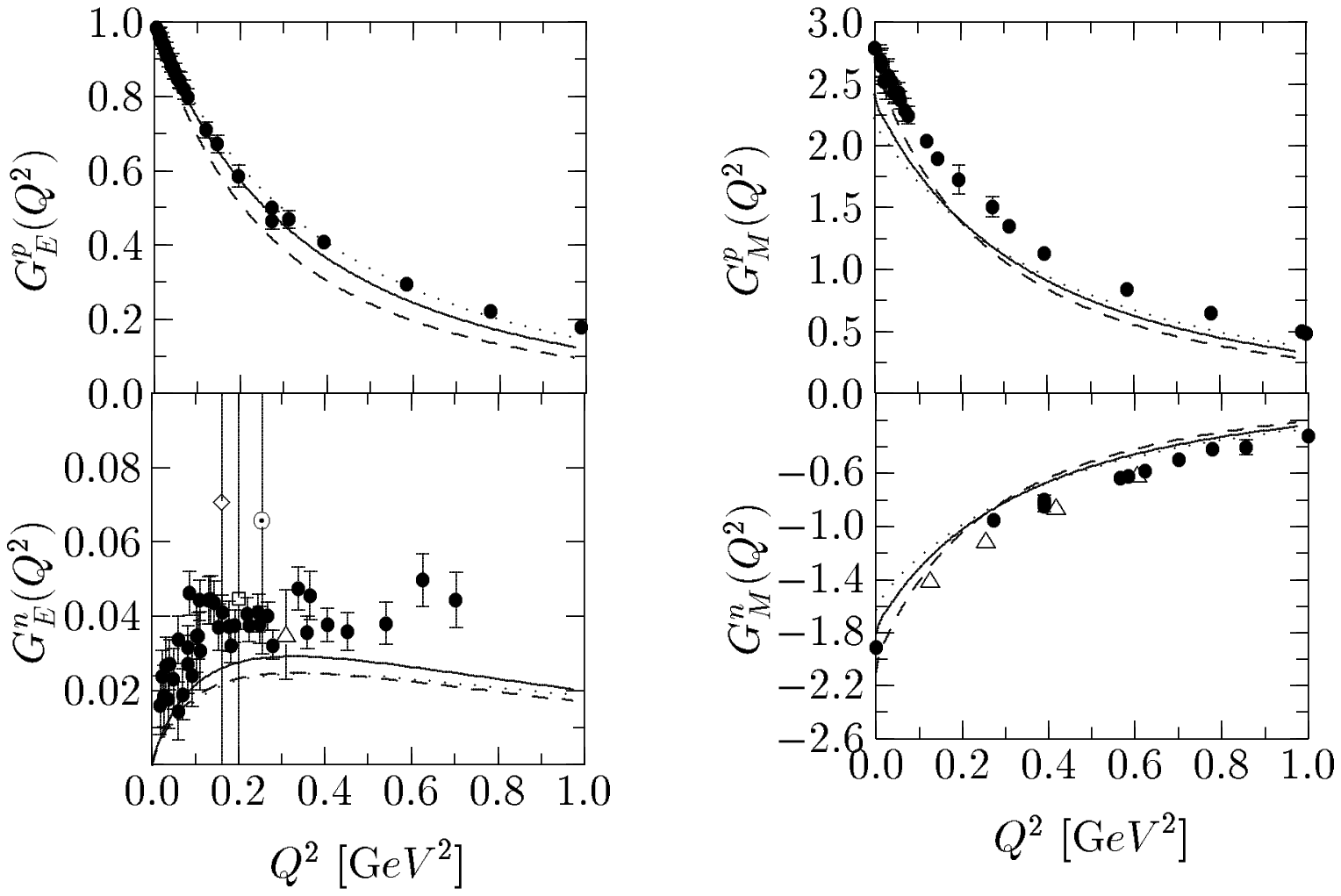}}\vskip4pt
\begin{center}
\parbox{14cm}{\footnotesize {\bf Fig. 1}:The electromagnetic form factors of
the nucleon as functions
of $Q^2$: In the left panel the electric form factors of the
nucleon are drawn, whereas in the right panel its magnetic formfactors
are depicted. The dashed curve corresponds to
the constituent quark mass $M=370 \mbox{MeV}$, while
solid curve is for $M=420\mbox{MeV}$.  The dotted curve
displays the case of $M=450\mbox{MeV}$.
The empirical data for the electric form factor of the proton
are taken from H\"ohler {\em et al.}\cite{holetal}, while those for
the proton one are from Platchkov {\em et al.}~\cite{pl}.
The other four points in the $G^{n}_{E}(Q^2)$ are results for $G^{n}_{E}$
extracted by Woodward {\em et al.}\cite{elecn1} (open diamond),
by Thompson {\em et al.}\cite{elecn2} (open box),
by Eden {\em et al.}\cite{edenetal} (open circle) and
by Meyerhoff {\em et al.}\cite{meyer} (open triangle).
The empirical data for the magnetic form factors are taken from
H\"ohler {\em et al.}~\cite{holetal} while
the data with open triangles are due to the most recent
experiment~\cite{Bruinsetal}.}
\end{center}

In this talk, we will present a recent investigation of
the electromagnetic properties of the SU(3) octet baryons in
this model.  Their explicit calculation can be found in
ref.\cite{Kimetal1,Kimetal2}.

Fig.~1 shows the electromagnetic form factors of the nucleon with the
constituent quark mass varied from 370 MeV to 450 MeV.
We select $M=420\;\mbox{MeV}$ for our best fit
consistently with the mass splittings.
It is found that the proton electric form factor is insensitive
to the constituent quark mass and it agrees with the empirical data by
H\"ohler (H\" ohler {\em et al.} 1990).

As can be seen from fig.2,
the result of the neutron electric form factors in SU(3) turns out
to be quite different from that in SU(2).  This discrepancy can be
explained by the fact that the isoscalar part of the charge operator
$\hat{Q}$ in SU(3) is not invariant under the rotation on the contrary
to SU(2).  This causes the reduction of the electric isoscalar form factors
in SU(3) and as a result the sizably smaller electric form factor
of the neutron is obtained, compared to that of the SU(2) model.
The nucleon magnetic form factors are displayed also in the right panel
of fig.1.

\vspace{0.8cm}
\centerline{\epsfysize=2.7in\epsffile{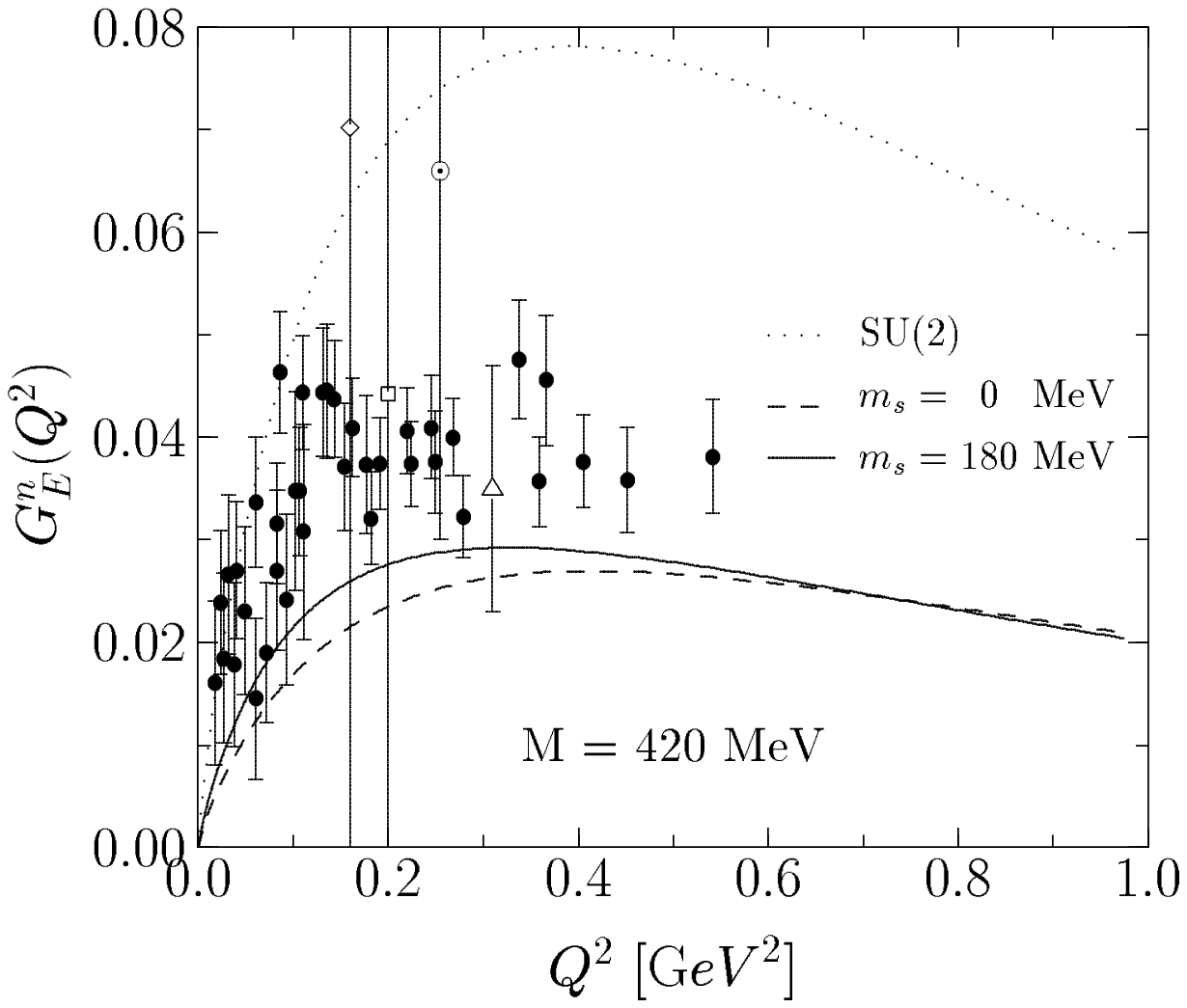}}\vskip4pt
\begin{center} \parbox{14cm}{\footnotesize{\bf Fig. 2}:
The neutron electric formfactor as a function of $Q^2$:
The solid curve corresponds to
the strange quark mass $m_s=180 \;\mbox{MeV}$, while
dashed curve draws without $m_s$.  The dotted curve
displays the case of the SU(2) model.
$M=420 \; \mbox{MeV}$ is chosen for the constituent quark mass.
The empirical and experimental data label the same cases as in
$G^{n}_{E}$ of fig. 1.}
\end{center}

\noindent They are also in a good agreement with experimental data.

In table~\ref{emff1} we list the magnetic moments and electromagnetic
charge radii of the SU(3) octet baryons.  Recently, Bae and McGovern
\cite{BaeMcGovern} made a $\chi^2$ analysis of the
hyperon magnetic moments for possible
hedgehog models, according to which
the present NJL model emerges as the best of hedgehog models.
The results are in a good agreement with the experimental
data within about $15\%$ which is more or less the upper limit
attained in any model with `` {\em hedgehog symmetry}''
\cite{Kimetal2}.
\begin{table}[h]
\caption{The electromagnetic static properties of the SU(3) octet baryons.
The constituent quark mass $M$ is used. }
\label{emff1}
\begin{tabular}{|c||cc|cc|cc|} \hline
$\mbox{Baryons}$ &
$\langle r^2\rangle_E[\mbox{fm}^2]$ & exp. &
$\mu_B[n.m.]$ & exp. &
$\langle r^2\rangle_M[\mbox{fm}^2]$ & exp. \\
\hline
$p$          & $~0.78$ & $~0.74$ & $~2.39$ & $~2.79$ & $0.70$ & $0.74$ \\
$n$          & $-0.09$ & $-0.11$ & $-1.76$ & $-1.91$ & $0.78$ & $0.77$ \\
$\Lambda$    & $-0.04$ & --      & $-0.77$ & $-0.61$ & $0.70$ & --     \\
$\Sigma^{+}$ & $~0.79$ & --      & $~2.42$ & $~2.46$ & $0.71$ & --     \\
$\Sigma^{0}$ & $~0.02$ & --      & $~0.75$ & --      & $0.70$ & --     \\
$\Sigma^{-}$ & $-0.75$ & --      & $-0.92$ & $-1.16$ & $0.74$ & --     \\
$\Xi^{0}$    & $-0.06$ & --      & $-1.64$ & $-1.25$ & $0.75$ & --     \\
$\Xi^{-}$    & $-0.72$ & --      & $-0.68$ & $-0.65$ & $0.51$ & --     \\
\hline
\end{tabular}
\end{table}
We now turn our attention to the electromagnetic form factors
of other SU(3) hyperons.
In fig. 3 we present the electric form factors for the
SU(3) octet hyperons while in fig. 4 we draw the magnetic form factors
of the hyperons.

\vspace{0.8cm}
\centerline{\epsfysize=2.7in\epsffile{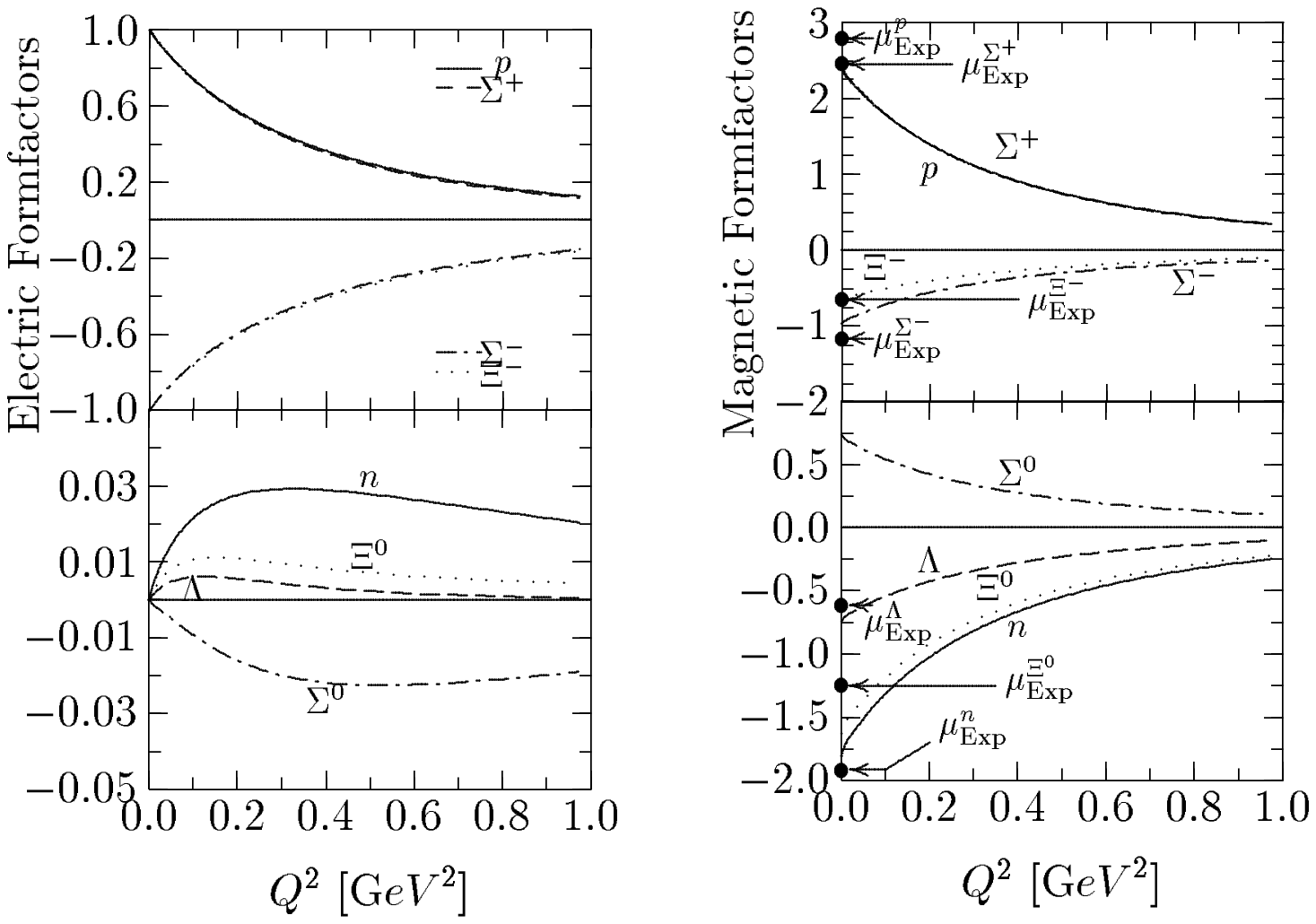}}\vskip4pt
\begin{center} \parbox{14cm}{\footnotesize {\bf Fig. 3}:
The electromagnetic formfactors of SU(3) octet hyperons:
In the left panel the corresponding electric form factors are drawn
and in the right panel the magnetic ones are depicted.}
\end{center}

This work has partly been supported by the BMFT, the DFG
and the COSY--Project (J\" ulich).  The work of M.V.P. is supported
in part by grant INTAS-93-0283.  A.B. would like to thank
the {\it Alexander von Humboldt Foundation} for a Feodor Lynen grant.

\end{document}